\documentclass[notitlepage,aps,pra,superscriptaddress]{revtex4-1}

\usepackage{color}
\usepackage{graphicx}  
\expandafter\let\csname equation*\endcsname\relax
\expandafter\let\csname endequation*\endcsname\relax
\usepackage{amsmath}
\usepackage{amssymb}
\usepackage[utf8x]{inputenc}
\usepackage{textcomp}
\usepackage{hyperref}
\usepackage{xcolor}
\hypersetup{
	colorlinks,
	linkcolor={blue!80!black},
	citecolor={red!70!black},
	urlcolor={blue!80!black}
}
\usepackage{amsthm}
\usepackage{bbm}
\usepackage{mathtools}
\mathtoolsset{centercolon}

\def\seti{\mathbf A}
\def\setii{\mathbf B}

\def\ket #1{\vert#1\rangle}

\def\six{\mathop{\mathrm s\mkern-1mu\mathrm i}\nolimits}
\DeclareRobustCommand{\sixnp}{\mathop{\mathrm s\mkern-1mu \textup \i}\nolimits}
\DeclareRobustCommand{\sixR}{\mathop{\hbox{\raisebox{.45em}{$\scriptstyle\rightharpoonup$}}\hspace{-.9em}\sixnp}}

\def\ch{\gamma}
\def\idty{\mathbbm 1}

\begin{document}

\title[Eigenvalue Measurement in Quantum Walks]{Eigenvalue Measurement of Topologically Protected Edge states in Split-Step Quantum Walks}

\author{Thomas~Nitsche}
\email{tnitsche@mail.upb.de}
\affiliation{Applied Physics, University of Paderborn, Warburger Straße 100, 33098 Paderborn, Germany}

\author{Tobias~Geib}
\affiliation{Institut für Theoretische Physik, Leibniz Universität Hannover, Appelstr. 2, 30167 Hannover, Germany}

\author{Christoph~Stahl}
\affiliation{Institut für Theoretische Physik, Leibniz Universität Hannover, Appelstr. 2, 30167 Hannover, Germany}

\author{Lennart~Lorz}
\affiliation{Applied Physics, University of Paderborn, Warburger Straße 100, 33098 Paderborn, Germany}

\author{Christopher~Cedzich}
\affiliation{Institut für Theoretische Physik, Leibniz Universität Hannover, Appelstr. 2, 30167 Hannover, Germany}
\affiliation{Institut für Theoretische Physik, Universität zu Köln, Zülpicher Straße 77, 50937 Köln, Germany}

\author{Sonja~Barkhofen}
\affiliation{Applied Physics, University of Paderborn, Warburger Straße 100, 33098 Paderborn, Germany}

\author{Reinhard~F.~Werner}
\affiliation{Institut für Theoretische Physik, Leibniz Universität Hannover, Appelstr. 2, 30167 Hannover, Germany}

\author{Christine~Silberhorn}
\affiliation{Applied Physics, University of Paderborn, Warburger Straße 100, 33098 Paderborn, Germany}

\begin{abstract}
We study topological phenomena of quantum walks by implementing a novel protocol that extends the range of accessible properties to the eigenvalues of the walk operator. To this end, we experimentally realise for the first time a split-step quantum walk with decoupling, which allows for investigating the effect of a bulk-boundary while realising only a single bulk configuration.
We approximate the symmetry protected edge states with high similarities and read out the phase relative to a reference for all modes.
In this way we observe eigenvalues which are distinguished by the presence or absence of sign flips between steps. Furthermore, the results show that investigating a bulk-boundary with a single bulk is experimentally feasible when decoupling the walk beforehand.
 
\end{abstract}

\maketitle

\section{Introduction}

Phenomena such as the quantum Hall effect \cite{klitzing_new_1980} and topological insulators \cite{kane_z_2_2005,konig_quantum_2007} aroused vivid interest in the study of the topological properties of physical systems. While these effects have been observed in semiconductor systems, experimental studies have been conducted on systems such as ultra cold atoms \cite{aidelsburger_measuring_2015, mancini_observation_2015, stuhl_visualizing_2015, groh_robustness_2016}, photonic model systems \cite{hafezi_robust_2011,kitagawa_observation_2012, rechtsman_photonic_2013}, solid-state systems \cite{xiong_evidence_2015, gooth_experimental_2017}, superconducting circuits \cite{flurin_observing_2017}, mechanical oscillators \cite{susstrunk_observation_2015} and microwave networks \cite{wang_observation_2009,hu_measurement_2015, poli_selective_2015}. 
In photonic systems, topological phenomena  can be accessed by implementing a split-step quantum walk on a 1D optical lattice \cite{kitagawa_exploring_2010, asboth_symmetries_2012, barkhofen_measuring_2017}.

The concept of quantum walks extends the model of classical random walks by taking the effects of interference and an internal degree of freedom of the walker into consideration. Possible experimental implementations include nuclear magnetic resonance \cite{du_experimental_2003, ryan_experimental_2005}, trapped ions \cite{schmitz_quantum_2009,zahringer_realization_2010} and atoms \cite{karski_quantum_2009, genske_electric_2013}. Considering photonic systems, translating the walker in the spatial degree of freedom might seem most straight-forward \cite{perets_realization_2008,peruzzo_quantum_2010,broome_discrete_2010,owens_two-photon_2011,sansoni_two-particle_2012,di_giuseppe_einstein-podolsky-rosen_2013,poulios_quantum_2014,xue_observation_2014}, however, implementations utilising time as the external degree of freedom outperform spatiallly multiplexed systems in terms of resource efficiency and stability \cite{  schreiber_photons_2010,schreiber_decoherence_2011,regensburger_photon_2011,schreiber_2d_2012,jeong_experimental_2013,boutari_large_2016,he_time-bin-encoded_2017}. Other possible degrees of freedom include spectral distributions or orbital angular momentum \cite{bouwmeester_optical_1999,cardano_quantum_2015}.

So far, the experimental investigation of topological phenomena has been focused on the demonstration of edge states, i.e. eigenstates of the systems, while eigenvalues of the walk operator have not been measured.
Employing a phase-reference method, in which we selectively interfere components of walker's wavefunction assigned to a certain step, position and polarisation with a reference of well-controlled phase, we are for the first time able to measure not only the intensities related to the eigenfunctions, but also the signs related to the eigenvalues of the walk operator.
This work also constitutes the first experimental implementation of a split-step quantum walk with decoupling as proposed in \cite{cedzich_bulk-edge_2016, cedzich_topological_2018}, which allows for investigating the effect of a bulk-boundary while realising only a single bulk configuration. Such a scheme thus not only requires a smaller set of different coin angles that have to be implemented, but also reduces the size of the space occupied by the walk, freeing positions which can now be used for routing of the phase-reference.
We implement this split-step quantum walk system exhibiting a bulk-boundary with decoupling by making use of a photonic platform \cite{schreiber_photons_2010} that allows for the read-out of the spatial and the coin degree of freedom as well as for dynamic coin operations \cite{nitsche_quantum_2016}. 

In our work, we build upon the comprehensive topological classification framework for infinite 1D lattices laid out in \cite{cedzich_topological_2018}. It predicts the emergence of symmetry protected edge states at the boundary of bulks with different symmetry indices. These symmetry protected eigenfunctions of the walks operator belong to the eigenvalues $\pm1$, which lie in the band gap of the system. Whereas the minimal number of these edge states is an invariant quantity, their associated eigenvalue ($+1$ or $-1$) may be changed under local perturbations and hence, the eigenvalues of edge states for half-chain quantum walks are not robust against local perturbation. We will investigate this behaviour for an experimental realization of the split-step quantum walk.
 
The article is structured as follows: We start with an introduction to the topological classification of 1D quantum walks in subsection \ref{subsec:topo_classification}, then present split-step quantum walks in subsection \ref{subsec:split_step_QW}, the relation between chiral symmetry and topological phases in subsection \ref{subsec:Chiral}, the concept of decoupling in subsection \ref{subsec:Decoupling}, the concrete settings in subsection \ref{subsec:Settings} as well as the expected eigenfunctions and eigenvalues in subsection \ref{subsec:Eigenfunctions}.
We then turn to the experimental concepts, namely the physical implementation in subsection \ref{subsec:Implementation}, the eigenstate distillation in subsection \ref{subsec:state_distillation} and the phase-reference method in subsection \ref{subsec:phase_reference}. 
The results concerning the evolution of the wavefunction and the validation of the eigenvalues are presented in subsection \ref{subsec:Wavefunction} resp. \ref{subsec:exp_eigenvalues}.
Finally, the conclusion is drawn in section \ref{sec:conclusion}.

\section{Theoretical Background}
\label{sec:background}

\subsection{Topological classification of one-dimensional Quantum Walks}
\label{subsec:topo_classification}

In this section we provide a rough sketch of the complete topological classification of symmetric quantum walks in one spatial dimension given in \cite{cedzich_topological_2018}. The setting of \cite{cedzich_topological_2018} is more general than actually needed for the purpose of this article, and we will narrow it down to the concrete setup as implemented in the experiment in the following section.

The systems under consideration are single particles with an internal degree of freedom evolving in discrete time on a lattice. Such systems are called \emph{discrete-time quantum walks}.
We here focus on the spatially one-dimensional case, for which the underlying Hilbert space is
\begin{equation}
\mathcal H=\bigoplus_{x\in\mathbb Z}\mathcal H_x,
\end{equation}
with finite dimensional \emph{cells} $\mathcal H_x$. 
On this Hilbert space, quantum walks are defined abstractly as unitary operators $W$ satisfying a locality condition which in this paper corresponds to a finite jump length $L<\infty$, i.e.
\begin{equation}
\langle\psi_x\vert W\psi_y\rangle=0\quad\text{for all }\psi_x\in\mathcal H_x,\psi_y\in\mathcal H_y\text{ with }  \vert x-y\vert>L.
\end{equation}
The general theory in \cite{cedzich_topological_2018} allows for more general notions of locality, but in the models considered below we even restrict to next-neighbour walks, i.e. $L=1$.

In each cell we consider the action of a group of discrete, involutive symmetries. By Wigner's theorem such symmetries are implemented by unitary or anti-unitary operators squaring to a phase times identity, which by standard arguments can be fixed to $\pm1$. The symmetries are further distinguished by their action on ``symmetric operators'', which in the present case are unitary operators satisfying predetermined relations with the symmetries. Concretely, we restrict our considerations to the following:

\begin{itemize}
	\item[] \textbf{chiral symmetry:} $\gamma W\gamma^\dagger=W^\dagger$, with $\gamma$ unitary,
	\item[] \textbf{particle-hole symmetry:} $\eta W\eta^\dagger=W$, with $\eta$ anti-unitary,
	\item[] \textbf{time-reversal symmetry:} $\tau W\tau^\dagger=W^\dagger$, with $\tau$ anti-unitary.
\end{itemize}

The groups generated by these symmetries either contain only one or all three, since any two of them multiply to the third. For each of these groups, the squares and the (anti-) unitarity together with the relations with symmetric operators constitute one of the symmetry types of the tenfold way \cite{altland_nonstandard_1997,kitaev_periodic_2009,cedzich_topological_2018}.

The eigenspaces at $\pm1$ of a symmetric unitary operator are invariant under the action of the symmetries. Similar to the Hamiltonian setting, we assume the bulk systems under consideration to be gapped at these symmetry-invariant points. Consequently, we call a quantum walk \emph{admissible} for a symmetry type if it is symmetric for the corresponding symmetry group and its bulk is gapped at $\pm1$.

The topological classification of admissible quantum walks is concerned with the question which admissible quantum walks 
may be continuously connected without breaking either the symmetries, the gap condition or locality.
A related question is which walks may be transformed into each other by perturbing one of them locally. Different from the continuous time case, in the unitary discrete time setting there are local perturbations that cannot be contracted, which is due to the extra symmetry invariant point in the spectrum at $-1$ \cite{cedzich_topological_2018}.
The complete classification in \cite{cedzich_topological_2018} answers these questions by assigning the three \emph{symmetry indices} $\six_-,\six_+$, and $\sixR$ to each admissible quantum walk. Here, $\six_\pm$ characterize the $\pm1$ eigenspaces and its values cannot be changed by continuous deformations that respect the symmetry. On the other hand, the \emph{right symmetry index} $\sixR$ characterizes the evolution asymptotically far on the right half-chain. It is invariant under continuous deformations, but in contrast to $\six_\pm$ also cannot be changed by locally perturbing the system.

An important physical consequence of this topological classification is 
the so-called \emph{bulk-boundary correspondence} for quantum walks: whenever two translation invariant admissible walks are joined spatially, the absolute value of the difference of their right symmetry indices $\sixR$ is a lower bound for the number of symmetry protected edge states which appear near the interface region. These edge states belong to the eigenvalues $\pm1$ and are therefore stable against continuous deformations that respect the symmetries. Since these symmetry protected edge states are the only eigenfunctions observed in the concrete examples below, we occasionally refer to them simply as eigenfunctions.

In order to provide experimental evidence for the emergence of symmetry protected edge states one typically measures the position distribution of a time evolved initial state. 
This is, however, not sufficient to reveal their topological nature since the emergence of localized states near local perturbations is a typical phenomenon \cite{ashcroft_mermin,ahlbrecht_molecular_2012}. Instead, in this publication we demonstrate an eigenvalue measurement on edge states of a topological quantum walk, to give evidence that the observed states near the boundary are indeed symmetry protected.

The appearance of symmetry protected edge states does not depend on how the crossover between two admissible systems is designed. An especially simple scenario where such edge states emerge is given by decoupling a given walk operator. This corresponds to a locally perturbed walk $W'=W_L\oplus W_R$ which has zero transition amplitudes between the right and left half chain of the cut.
The right symmetry index $\sixR(W)=\six(W_R)$ may then be calculated as the combined symmetry indices $\six_-$ and $\six_+$ of the right half-chain walk: $\sixR(W)=\six_+(W_R)+\six_-(W_R)$. The existence of such decoupling is guaranteed by the \emph{gentle decoupling theorem}, and this decoupling can even be chosen to be a continuous perturbation \cite{cedzich_topological_2018}. 

However, whether the symmetry protected edge states correspond to eigenvalues at $+1$ or $-1$ depends on the crossover, since in contrast to their sum $\sixR(W)$, the indices $\sixR_\pm(W)=\six_\pm(W_R)$ themselves are not stable under local perturbations. Exactly this dependence of $\sixR_\pm(W)$ on the crossover will be analysed and experimentally validated in the present publication by choosing different decouplings.

\subsection{Split-Step Quantum Walks}
\label{subsec:split_step_QW}
While the above definition of quantum walks is the appropriate setting for the general topological classification in \cite{cedzich_topological_2018}, there is a particular way to construct quantum walks which comes in handy for experimental realizations: like for classical random walks, for \emph{coined quantum walks} the walker's path is determined by a coin toss. However, while in the classical case the walker takes one of the possible paths, in the quantum setting the amplitude of the wave function is split up into components taking the different paths.
This results in the superposition of components originating from different paths and consequently interference effects are observed.

When all local Hilbert spaces $\mathcal H_x$ are equal, the state ${| \psi \rangle}$ of the quantum walker is an element of the tensor product space $\ell^2(\mathbb Z)\otimes\mathcal{H}_c$ of the position Hilbert space $\ell^2(\mathbbm Z)$ and the coin Hilbert space $\mathcal H_c$.
Denoting by $| x \rangle $ and $| i \rangle $ the basis vectors of the position and the coin space, respectively, the walker's wave function at time $t$ is 

\begin{equation}
| \psi (t) \rangle =  \sum_{x,i} \psi_{x,i}(t)| x ,\, i\rangle ~.
\label{eq:Psi}
\end{equation}
with the time-dependent amplitudes $\psi_{x,i}(t) \in \mathbb C$. Its evolution in discrete time steps is determined by applying the quantum walk operator $W$. Note that here and in the entire section \ref{sec:background} we omit the operator hats for readability.
\begin{equation}
|\psi(t+1) \rangle = W |\psi(t) \rangle.
\label{eq:SC}
\end{equation}

A coined quantum walk $W$ consists of a sequence of coin operations $C$, the analog of classical coin tosses, acting exclusively on the internal state and shift operations $S_j$ (and $S_j^\dagger$), modifying only the positional degree of freedom:

\begin{equation}
C \ket {x,\, i}=\sum_j C_{ij}(x)\ket {x,\, j} \quad\text{ and }\quad S_j\ket {x,\,i}=\ket{x+\delta_{ij},\,i},
\end{equation}
where $S_j$ ($S_j^\dagger$) shifts components in the internal degree of freedom indexed $j$ to the right (left) while leaving the other internal degrees of freedom unchanged.

In our realisation of a quantum walk the internal state $\ket{\psi}\in\mathcal H_c=\mathbb{C}^2$ is represented in terms of the polarisation basis vectors $\ket{H}=(1,0)^{\mathrm T}$ and $\ket{V}=(0,1)^{\mathrm T}$.

The split-step quantum walk is a specific realisation of a coined quantum walk. First introduced in \cite{kitagawa_exploring_2010} as a simple example of a symmetric quantum walk, it is defined by
\begin{equation}\label{eq:walk_splitstep}
W(\theta_1,\theta_2)=S_\downarrow C(\theta_1)S_\uparrow C(\theta_2),
\end{equation}
where $S_\uparrow=S_H$ and $S_\downarrow=S_V^\dagger$ shift the $| H \rangle$ state to the right and the $\ket V$ state to the left, respectively, and
\begin{equation}\label{eq:coin}
C(\theta)=e^{-i\theta\sigma_x}=\begin{pmatrix} \cos(\theta) &   -i\sin(\theta)   \\  -i\sin(\theta)   &   \cos(\theta) \end{pmatrix}
\end{equation}
rotates the internal degree of freedom by an angle $\theta$ around the $\sigma_x$-axis. Note that the split-step walk is typically defined with rotations around the $\sigma_y$-axis. Since these two implementations are unitarily equivalent and in the experiment described below we implement the $\sigma_x$-rotation, we take \eqref{eq:walk} with coin \eqref{eq:coin} as the definition of the split-step walk. The walk is of symmetry type $\mathrm{BDI}$, i.e. it has chiral, time-reversal and particle-hole symmetry with each of them squaring to the identity. It exhibits a rich structure of topological phases, and therefore has become the working example in many publications concerning topological effects in quantum walks \cite{kitagawa_exploring_2010,kitagawa_observation_2012,kitagawa_topological_2012,asboth_symmetries_2012,cedzich_complete_2018}. The phase diagram in Figure~\ref{fig:harlequin} shows the different topological phases of the split-step quantum walk.
The interactive web tool in \cite{sse} allows the user to explore how the eigenfunctions and the symmetry indices of the split-step walk change with modifications of the parameters $\theta_1$ and $\theta_2$ as well as for different decouplings.

In the experiment described below the walk we implement is not the split-step walk but the related model
\begin{equation}\label{eq:walk}
W=SC(\theta_1)SC(\theta_2),
\end{equation}
where $S=S_H S_V^\dagger$ denotes the bidirectional conditional shift. 
Since this walk protocol contains two shift operations, a walker which is initially localized on even (odd) positions, never leaves the even (odd) sub-lattice. Thus, on each of these sublattices the walk \eqref{eq:walk} implements effectively a split-step walk with doubled jump length. Since our initial states will always be localized at $x=0$, we restrict considerations to the even sub-lattice from now on. Rescaling this even sub-lattice by $x\mapsto x/2$ the effective walk is just described by \eqref{eq:walk_splitstep}.

\subsection{Chiral symmetry and topological phases}
\begin{figure}[]
	\centering
	\includegraphics[width=0.75\textwidth]{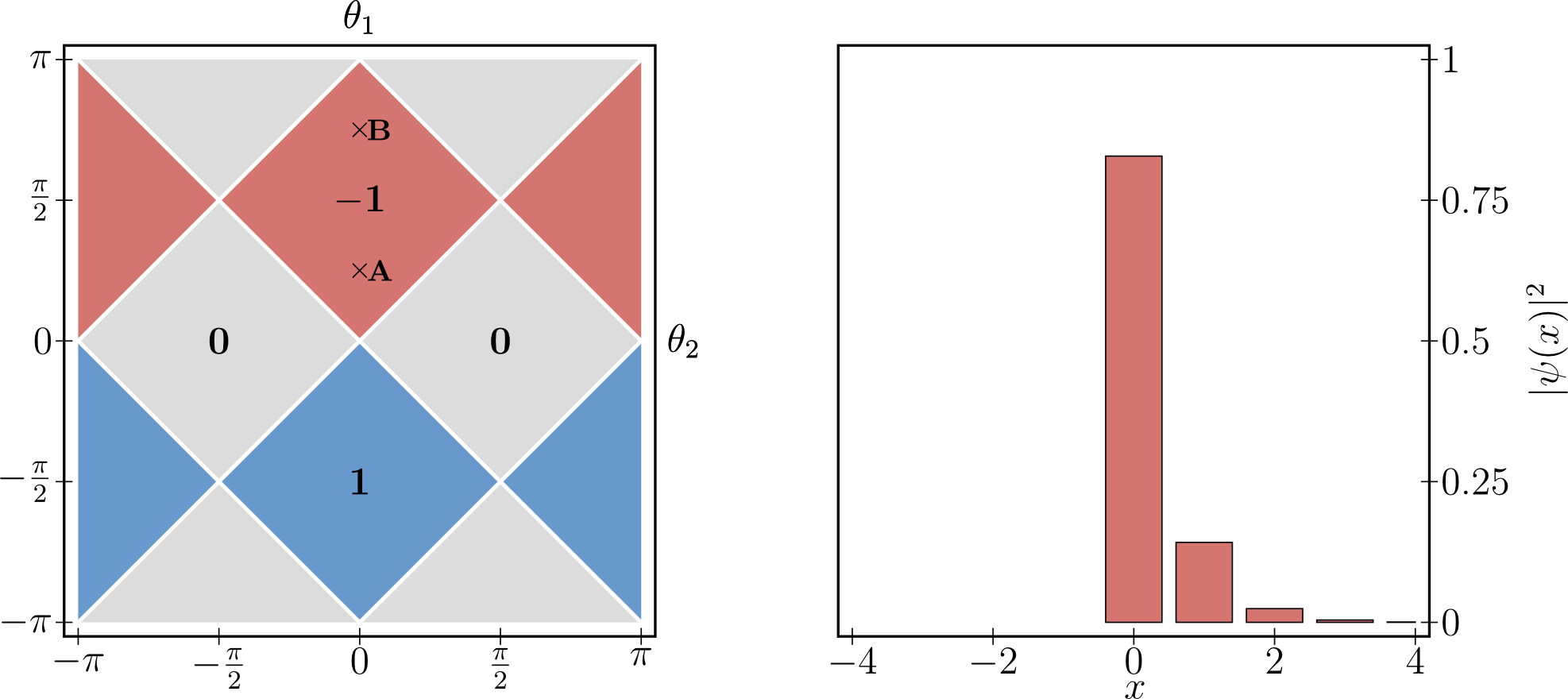}
	\caption{Parameter plane for the split-step walk \eqref{eq:walk_splitstep} with regions of constant index (left). The crosses mark the parameters for our two settings ($\seti$/$\setii$).\newline
		Intensity distribution $|\psi(x)|^2$ of the corresponding eigenfunction (right). The distribution is the same for both settings.}
	\label{fig:harlequin}
\end{figure}

\label{subsec:Chiral}
Being of symmetry type $\mathrm{BDI}$ the split-step walk has chiral symmetry, i.e.\ there exists a unitary $\ch$ with $\ch^2=+\idty$ such that $\ch W\ch^\dagger=W^\dagger$. This symmetry operator is given by
\begin{equation}\label{eq:gamma}
\ch=\bigoplus_{x\in\mathbb Z}\ch_0,\quad\text{with}\quad\ch_0=\begin{pmatrix}
-\sin(\theta_2) & -i\cos(\theta_2)\\
i\cos(\theta_2) & \sin(\theta_2)
\end{pmatrix}.
\end{equation}
Note that the $\theta_2$-dependence of $\ch$ is due to the choice of basis: conjugating with $C(\theta_2/2)$ gives $C(\theta_2/2)^\dagger\ch_0C(\theta_2/2)=\sigma_y$. 

For translation invariant chiral symmetric walks with $\ch^2=+\idty$, the symmetry index $\sixR$ is given by the winding number of the determinant of the upper right block of the walk in the chiral eigenbasis, considered as a complex function $\mathbb T\ni k\mapsto c(k)$ on momentum space \cite{asboth_symmetries_2012,cedzich_complete_2018}. For the split-step walk this function is given by
\begin{equation}
c(k)=\sin(\theta_1)\cos(\theta_2)+\cos(\theta_1)\sin(\theta_2)\cos(k)-i\cos(\theta_1)\sin(k),
\end{equation}
and the corresponding values of the index are shown in the parameter plane in Fig.~\ref{fig:harlequin}.

\subsection{Decoupling}
\label{subsec:Decoupling}

To construct a decoupling for the split-step walk we make use of its special form. Replacing in \eqref{eq:walk_splitstep} the local coin $C(\theta_1)$ at $x=0$ by a reflective coin, i.e.\ a coin whose diagonal elements are zero, decouples the resulting walk between $x=0$ and $x=-1$. To reproduce this decoupled split-step setting as an effective sublattice walk of the experimentally accessible walk, we need to replace the coin $C(\theta_1)$ at $x=-1$ in \eqref{eq:walk}.

The decoupling coins within the parameter regions are the rotations by angles $\theta=\pm \pi/2$, for which \eqref{eq:coin} indeed has only zeros on the diagonal. Whenever a walk in a non-trivial phase, i.e. with non-vanishing $\sixR$ is decoupled, exponentially localized eigenstates are predicted at the interface by bulk-boundary correspondence. As remarked above, however, it depends on the specific decoupling whether these eigenfunctions correspond to the eigenvalue $+1$ or $-1$. These two possibilities can be distinguished by an additional (walk specific) invariant, namely $\sixR_-(W)=\six_-(W_R)$ of the decoupled walk $W'=W_L\oplus W_R$.

\subsection{Settings}
\label{subsec:Settings}

\begin{figure}[h]
	\centering
	\includegraphics[scale=.75]{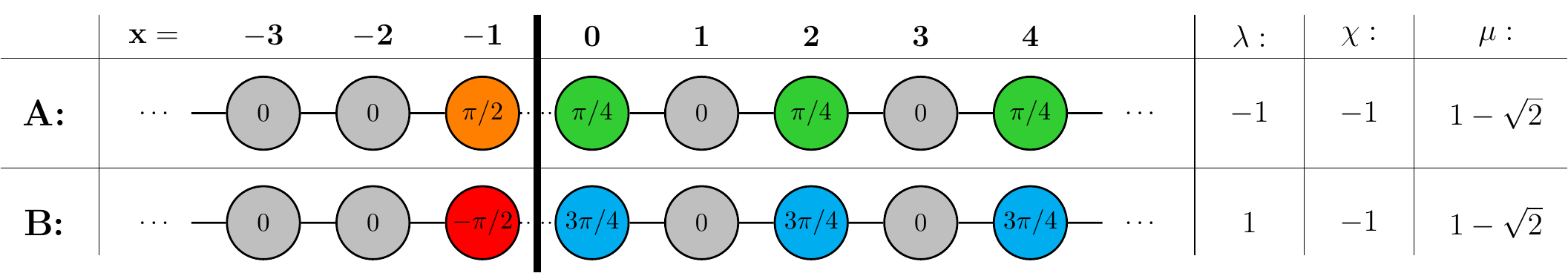}
	\caption{Schematic picture of the coin angle set-up for the walk \eqref{eq:walk} in our two settings. The coin angles at $x=-1$ decouple the walk between $x=-1$ and $x=0$. $\lambda$ denotes the eigenvalue, $\chi$ the chirality and $\mu$ the decay coefficient of the edge state, respectively.}
	\label{fig:coins}
\end{figure}

We consider two settings with different values for the coin. In both settings the effective walks have non-trivial symmetry index and are decoupled between $x=-1$ and $x=0$. The aim is to determine the eigenvalues of the eigenfunctions emerging to the right at the boundary, i.e.\ on $\mathcal H_R=\bigoplus_{x\geq0}\mathcal H_x$. The two settings for \eqref{eq:walk} are the following (see Figure \ref{fig:coins}):

\paragraph*{\bf{Setting $\seti$:}}
\begin{equation}\label{eq:coin_setting_one}
\theta_2=\pi/4\qquad\text{and}\qquad\theta_1(x)=\begin{cases}\pi/2 & x=-1\\0 &\text{else}\end{cases}
\end{equation}
\paragraph*{\bf{Setting $\setii$:}}
\begin{equation}\label{eq:coin_setting_two}
\theta_2=3\pi/4\qquad\text{and}\qquad\theta_1(x)=\begin{cases}-\pi/2 & x=-1\\0 &\text{else}\end{cases}
\end{equation}

In both cases $\theta_1$ is chosen to decouple the walk on the even sub-lattice. We do not need to specify the left half chain here, since we are only interested in edge states of the decoupled walk located to the right.
Hence, only the coin configuration for $x>0$ determines the phase of the effective  walk, and we infer from the phase diagram in Figure~\ref{fig:harlequin} that in both settings the corresponding symmetry index is $\sixR(W_{\seti/\setii})=-1$, which predicts the emergence of edge states.

\subsection{Eigenfunctions and Eigenvalues}
\label{subsec:Eigenfunctions}

To compute the eigenfunctions of $W$ in both settings, we first note that $W$ and $\ch$ commute on the $\pm1$-eigenspaces of $W$. Therefore, we can jointly diagonalize $W$ and $\ch$ on these eigenspaces. The (un-normalized) eigenvectors $\varphi_0^\chi$ of $\ch_0$ in \eqref{eq:gamma} are of the form
\begin{equation}\label{eq:eigenstate-structure}
\varphi^\chi_0=\begin{pmatrix}
i\cos(\theta_2)\\
-(\sin(\theta_2)+\chi)
\end{pmatrix},
\end{equation}
where $\chi\in\{+1,-1\}$ denotes the chirality, i.e.\ the eigenvalue of $\ch$. Since edge-states of decoupled translation invariant systems have to decay exponentially in the bulk \cite{cedzich_complete_2018} and we are only interested in eigenfunctions of the walk $W$ located to the right, we choose the following ansatz:

\begin{align}\label{eq:eigenstate}
\phi_\lambda^\chi(x) &= \begin{cases}0 & x<0\\a\cdot\varphi^\chi_0 & x=0\\ \mu^x\varphi^\chi_0 &x>0\end{cases}, \quad\text{fulfilling the eigenvalue equation} &&W\phi^\chi_\lambda=\lambda\phi^\chi_\lambda.
\end{align}
Here, the free parameter $a$ takes care of the boundary condition which is determined by the choice of the decoupling coin, and $\mu$ denotes the exponential decay rate away from the boundary. Note that $\phi^\chi_0$ is normalizable if $|\mu|<1$.
To solve for $\mu$, we evaluate the eigenvalue equation in the bulk, i.e.\ without taking the boundary into account. This leads to
\begin{align}
0 &=\lambda\cos(\theta_2)\mu-\chi\sin(\theta_2)-1\\
0 &=\chi\cos(\theta_2)\mu-\lambda\sin(\theta_2)-\chi\lambda,
\end{align}
which are solved by
\begin{align}
\mu(\lambda,\chi)=\frac{1+\chi\sin(\theta_2)}{\lambda\cos(\theta_2)}.
\end{align}
In both settings $\seti$ and $\setii$ with $\theta_2=\pi/4$ and $\theta_2=3\pi/4$, respectively, we have
\begin{align}
|\mu(-1,1)|=|\mu(1,1)|&=|1+\sqrt2|>1,\\
|\mu(-1,-1)|=|\mu(1,-1)|&=|1-\sqrt2|<1.\label{absmu}
\end{align}
Therefore, by the condition $|\mu|<1$ in our setting the eigenfunction located to the right of the bulk must have chirality $\chi=-1$.

In order to determine $a$ we have to take into account the boundary condition. Choosing the solutions of the eigenvalue equation with $\chi=-1$ 
leads to the following equations for $a$:
\begin{align}
0 &= a\cos(\theta_2)(\lambda+\sin(\theta_1))\\
0 &=(a-1)(1-\sin(\theta_2))\lambda.
\end{align}
Note that the first equation is actually independent of $a$ whenever $a\neq0$. Hence, the first equation rules out one of the two possibilities in \eqref{absmu}. We get $\lambda=-1$ for $\theta_1=\pi/2$ (setting $\seti$) and $\lambda=1$ for $\theta_1=-\pi/2$ (setting $\setii$). In both settings, however, we must have $a=1$ for the second equation to be satisfied.

The eigenfunctions of $W$ located to the right are thus given by
\begin{align}
\varphi_R(2x) = \begin{cases}0 & x<0\\ c (1-\sqrt2)^x\begin{pmatrix}
i\cos(\theta_2)\\
1-\sin(\theta_2)
\end{pmatrix} &x\geq0,\end{cases}
\label{eq:theo_eigenfunction}
\end{align}
with the normalization factor $c=\bigl((1+\sqrt2)(1-\sin(\theta_2)\bigr)^{-\frac{1}{2}}$.

\newpage
\section{Experimental Implementation}
\subsection{Time-multiplexing setup}
\label{subsec:Implementation}

The realisation of the two described settings and the direct measurement of the eigenvalues require a stable, flexible experimental platform, as a phase-stable evolution incorporating a dynamic coin operation has to be ensured over a sufficiently large number of steps. 
Our system of choice relies on a well-established time-multiplexed architecture utilising fibre loops for discrete-time quantum walks (DTQW) \cite{  schreiber_photons_2010,schreiber_decoherence_2011,schreiber_2d_2012}. The dynamic coin operation implemented with a fast-switching electro-optic modulator (EOM) makes it suitable for a wide range of experiments, including the investigation of topological phenomena \cite{nitsche_quantum_2016,barkhofen_measuring_2017}.

Previous photonic implementations allowed for accessing topological invariants associated with probability distributions or amplitudes within a certain step of the walk \cite{kitagawa_observation_2012, barkhofen_measuring_2017, cardano_detection_2017,zhan_detecting_2017, chen_observation_2018, wang_detecting_2018, xu_measuring_2018}. However, topological properties can also manifest themselves in the emergence of eigenstates with associated eigenvalues that are revealed by the phase relation between the walker's wavefunctions for two consecutive steps. So far, this phase relation has not been investigated experimentally. By interfering the walker with a reference of fixed phase, we are now able to probe this feature as well.

Our implementation relies on a photonic walker implemented by an attenuated coherent laser pulse with its polarisation representing the internal (coin) degree of freedom. In this way, our system makes use of the equivalence of coherent light and a single quantum particle when propagating in a linear optical network \cite{paul_introduction_2004}.

\begin{figure}[t]
	\centering 
	\includegraphics[width=0.9\columnwidth]{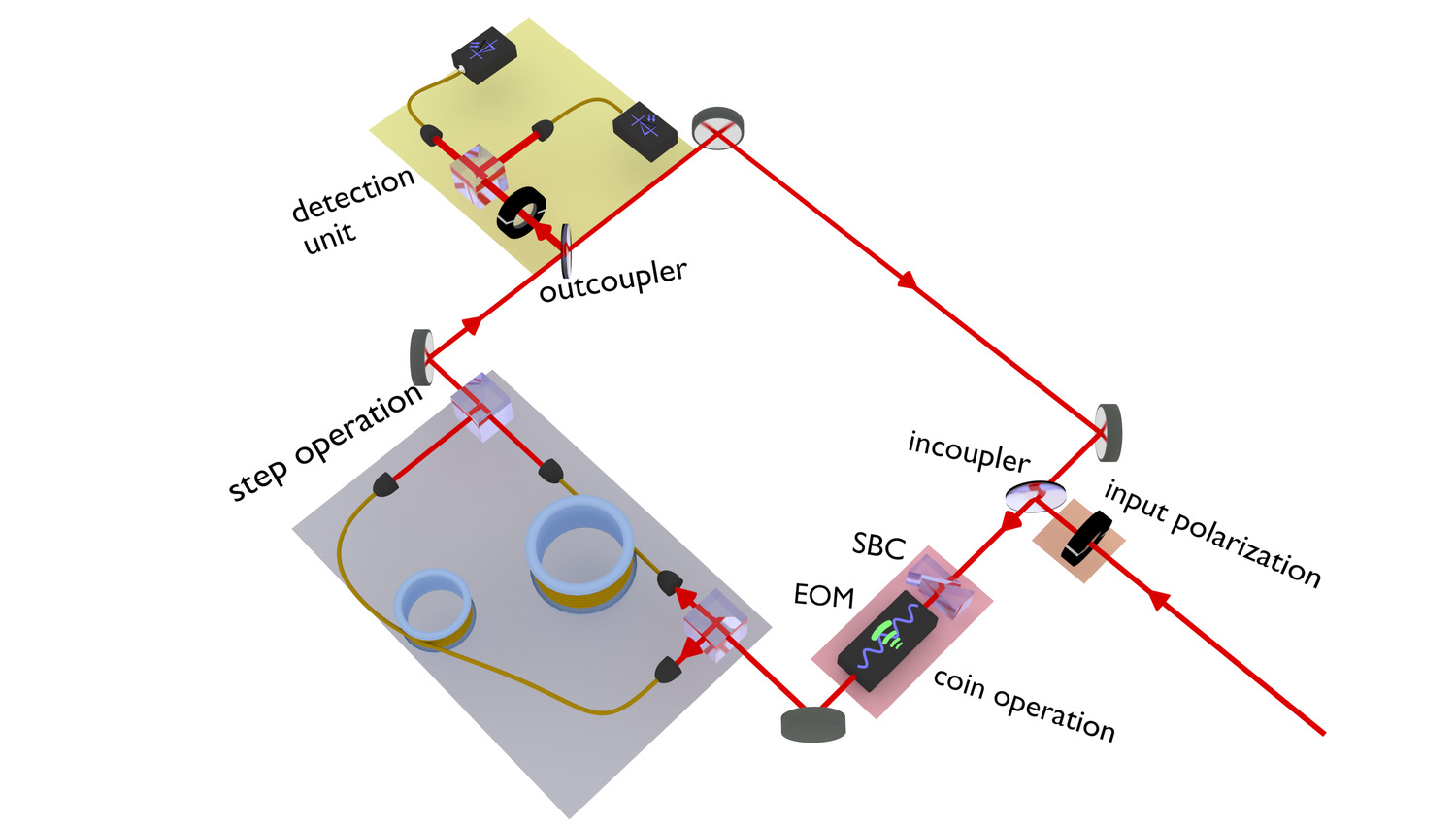}
	\caption{Schematic of our setup: The light is coupled in and out of the setup with probabilistic in- and outcouplers. The coin operation is carried out with a static Solei-Babinet compensator (SBC) and a dynamic electro-optic modulator (EOM). The step operation is carried out by splitting the light into two fibres of different length. The detection unit allows for polarisation-resolved read-out.}
	\label{fig:Setup}
\end{figure}

Figure \ref{fig:Setup} shows the physical implementation of the quantum walk setup:
Our time-multiplexing architecture relies on translating the external (position) degree of freedom of the walker into the time domain by splitting the pulses up spatially, routing them through fibres of different length and subsequently merging the two paths again. The ratio of this probabilistic splitting taking place at polarising beam splitters (PBS) is determined by the polarisation state of the walker. This internal degree of freedom is acted upon by coin operations implemented with static (Soleil-Babinet compensator, SBC) and dynamic (electro-optic modulator, EOM) polarisation optics. The remarkable characteristic of the EOM is that its switching speed allows for addressing individual positions within the walk. This dynamic coin enables us implementing an alternating coin needed for the split-step scheme as well as the reflecting and transmitting coin operations indispensable in directing the reference pulse along a certain path (see section \ref{subsec:phase_reference}). The polarisation degree of freedom can also be accessed in the read-out process, since our detection unit comprising another PBS and 2 avalanche photo-diodes (APDs) is polarisation-resolving.

\subsection{Eigenstate distillation}
\label{subsec:state_distillation}
Since the exponentially localised eigenstate cannot be excited directly with an initial state just occupying one position, we need a method to prepare an approximate eigenstate of the system. In order to do so, we make use of the spreading behaviour of the split-step quantum walk.
In systems for which the translational invariance is only broken at the edge, the eigenstates are exponentially localized \cite{cedzich_complete_2018}.  If we choose an initial state near a boundary and let the system evolve for some time, the components of the state which have no overlap with the localised state will propagate away from the boundary, while the content which overlaps will stay. Consequently, we concentrate our study onto the 3 positions near the boundary, which is justified due to the exponential decay of the state (see eq. \eqref{eq:theo_eigenfunction}). By renormalizing the remaining state, we prepare an approximate eigenfunction. 
For quantifying how closely we have approached the theoretically expected distribution, we use the similarity which is obtained by summing up the square roots of the products of the theoretical and experimental probabilities for the relevant positions within the step that is examined. It can possibly assume values between 0 (no overlap of intensities) and 1 (perfect overlap of intensities).

\begin{eqnarray}
d = \left| \sum_x \sqrt{P_{H,x}^{(\mathrm{theo})} \cdot P_{H,x}^{(\mathrm{exp})}}
+ \sum_x \sqrt{P_{V,x}^{(\mathrm{theo})} \cdot P_{V,x}^{(\mathrm{exp})}} \right|
\label{eq:similarity}
\end{eqnarray}

\subsection{Phase-Reference Method}
\label{subsec:phase_reference}
\begin{figure}[t]
	\centering 
	\includegraphics[width=10cm]{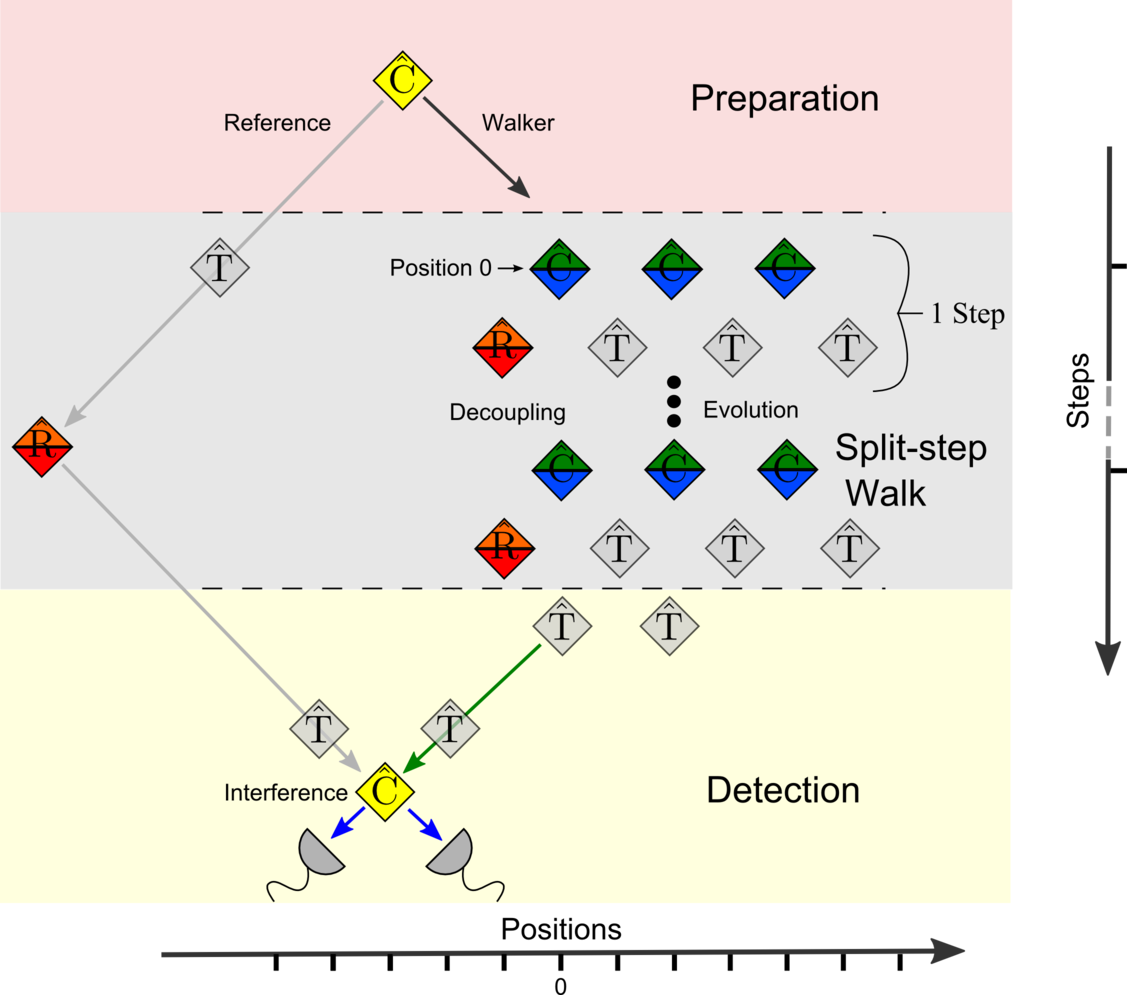}
	\caption{Schematic of our implementation of a split-step quantum walk followed by a phase readout via the interference method. Squares with $\hat{\mathrm{T}}$ denote a transmission operation, while $\hat{\mathrm{R}}$ refers to reflection and $\hat{\mathrm{C}}$ to mixing coins, for which the exact form can be different depending on setting and position.}
	\label{fig:Scheme}
\end{figure}

Figure \ref{fig:Scheme} illustrates how the dynamic coin operation can be harnessed to implement a split-step quantum walk followed by a phase readout via the interference method: By applying a mixing coin at the initial position (marked $\hat{C}$) the walker is split into a vertical component travelling through the shorter fibre, i.e. being translated to the left in the schematic and a horizontal component running through the longer fibre, which corresponds to a translation to the right. The part going to the left constitutes the reference. In order to prevent it from mixing with the split-step walk taking place on the right as well as from losing intensity to positions that are not on the desired path, the EOM switches to identity (marked $\hat{T}$ for transmission in Figure \ref{fig:Scheme}) on the positions where the reference is found.
For the phase read-out the reference needs to interfere with the light having undergone the walk, so the travelling direction is inverted by switching a reflection (marked $\hat{R}$ in Figure \ref{fig:Scheme}) in the middle of the propagation.

The light translated to the right in the initial splitting constitutes the input state of the split-step quantum walk with decoupling. The decoupling is realised by a reflection operation implemented on the decoupling position. The split-step walk (see \eqref{eq:coin_setting_one} and \eqref{eq:coin_setting_two}) incorporates a mixing coin (indicated by $\hat{C}$ in Figure \ref{fig:Scheme}) on even positions and the identity coin, corresponding to a transmission operation, on uneven positions.

The scheme shown in Figure \ref{fig:Scheme}, (a) brings the reference to interfere with the vertical light ending up at position 0 of the split-step walk. At the position where the light from the walk and reference interfere a mixing coin has to be applied. The ratio of the intensities at the two detection positions (marked by detectors) then allows obtaining information on the phase relation.
All polarisations, positions and steps can be accessed analogously, as long as the proper routing of the reference and the pulse under investigation can be ensured. The basic principle remains the same in all runs, but the actual positions of the switching differ. For this purpose, a flexible dynamic coin operation is indispensable.

We access two regimes with topologically different eigenvalues by applying a coin of either $\hat{C}_\mathrm{EV-1}= e^{-i \sigma_x \cdot \frac{\pi}{4}}$ for an eigenvalue of -1 (setting $\seti$) or $\hat{C}_\mathrm{EV+1}= e^{-i \sigma_x \cdot \frac{3}{4} \pi}$ for an eigenvalue of +1 (setting $\setii$) and $\hat{T}= e^{-i \sigma_x \cdot 0 \pi} = \idty$ for the transmission as well as $\hat{R}= e^{-i \sigma_x \cdot \frac{\pm \pi}{2}}$ for the reflection. All of these coins can be implemented with a combination of static SBC and dynamic EOM operations.

The interference of a certain component of the walker with the reference together with the application of a mixing coin (in this case the balanced Hadamard coin $\hat{C}_\mathrm{Had}$) results in the following expression for the wave function at the position where the interference takes place (labelled "Interference" in Figure \ref{fig:Scheme}):

\begin{equation}
|\Phi \rangle = \hat{C}_\mathrm{Had} \left( \begin{array}{c}
e^{i \alpha_\mathrm{w}} \sqrt{I_\mathrm{w}}\\
e^{i \alpha_\mathrm{r}} \sqrt{I_\mathrm{r}}
\end{array} \right)
=  \frac{1}{\sqrt{2}}\begin{pmatrix}
1 & 1\\
1 & -1
\end{pmatrix}
\left( \begin{array}{c}
e^{i \alpha_\mathrm{w}} \sqrt{I_\mathrm{w}}\\
e^{i \alpha_\mathrm{r}} \sqrt{I_\mathrm{r}}
\end{array} \right)
\label{eq:phi_meas}
\end{equation}

Here $\alpha_\mathrm{w}$ and $\alpha_\mathrm{r}$ denote the phase of the walker resp. of the reference. $\sqrt{I_\mathrm{w}}$ and $\sqrt{I_\mathrm{r}}$ are the square roots of the walker's and the reference's intensities.
The detected intensities $I_\mathrm{H}$ and $I_\mathrm{V}$ for the horizontal and the vertical detector are then given by the following expression:

\begin{equation}
\begin{split}
I_\mathrm{H} = \frac{1}{2} (I_\mathrm{w} + I_\mathrm{r} - 2 \sqrt{I_\mathrm{w}} \cdot \sqrt{I_\mathrm{r}} \sin(\alpha_\mathrm{r}-\alpha_\mathrm{w})) )\\
I_\mathrm{V} = \frac{1}{2} (I_\mathrm{w} + I_\mathrm{r} + 2 \sqrt{I_\mathrm{w}} \cdot \sqrt{I_\mathrm{r}} \sin(\alpha_\mathrm{r}-\alpha_\mathrm{w})) )
\end{split} 
\label{eq:I_H_I_V}
\end{equation}

Thus we can deduce the phase difference between the reference and a certain component of the walker from the measured intensities at the read-out positions (marked by a detector symbol in Figure \ref{fig:Scheme}):

\begin{equation}
M=\sin(\alpha_\mathrm{r}-\alpha_\mathrm{w})=\frac{I_\mathrm{V}-I_\mathrm{H}}{ 2 \sqrt{I_\mathrm{w}} \cdot \sqrt{I_\mathrm{r}}}
\label{eq:M}
\end{equation}

Note that the M-parameter given by this formula is $\sin(\alpha_\mathrm{r}-\alpha_\mathrm{w})$, which is not an injective function. However, it still provides a clear distinction between cases in which we expect an eigenvalue of 1 and cases with an eigenvalue of -1.

Distinguishing these eigenvalues requires monitoring how the M-parameter changes from step to step. For an eigenvalue of -1 (setting $\seti$), the walker's phase $\alpha_\mathrm{w}$ is expected to change by $\pi$, while for an eigenvalue of 1 (setting $\setii$) it is either 0 or integer multiples of $2 \pi$. Since the phase of the reference $\alpha_\mathrm{r}$ can be assumed to be constant for all numbers of steps, monitoring the step-wise evolution of the M-parameter for the two settings should clearly reveal the different eigenvalues. 
In the experiment, the eigenvalues will be manifested in the relative intensities of horizontal and vertical light at the read-out positions.
Accessing the eigenvalues requires the read-out of both polarisations for the three inner positions. Furthermore, the state of the walker over 3 steps (6,7 and 8) is monitored. Since each measurement just yields information for a certain position, polarisation and step, obtaining the full information makes 18 individual measurement runs necessary. Observing the differences between settings $\seti$ and $\setii$ doubles the number of measurements required, so that eventually 36 data sets have to be taken.

\section{Results} \label{sec:results}

As outlined previously, the experiment aims at accessing the evolution of eigenvalues over 3 consecutive steps. As we are limited by losses to numbers of physical roundtrips around 22, the actual state of the walker will constitute an approximation of the ideal eigenstate. We quantify the quality of the approximation by calculating the similarity between the experimental intensity and the ideal eigenstate according to \eqref{eq:similarity}. Furthermore, the phase-reference method indicates the eigenvalue of the system under investigation.

\subsection{Evolution of the Wave function}
\label{subsec:Wavefunction}

\begin{figure}[t]
	\centering 
	\includegraphics[width=\columnwidth]{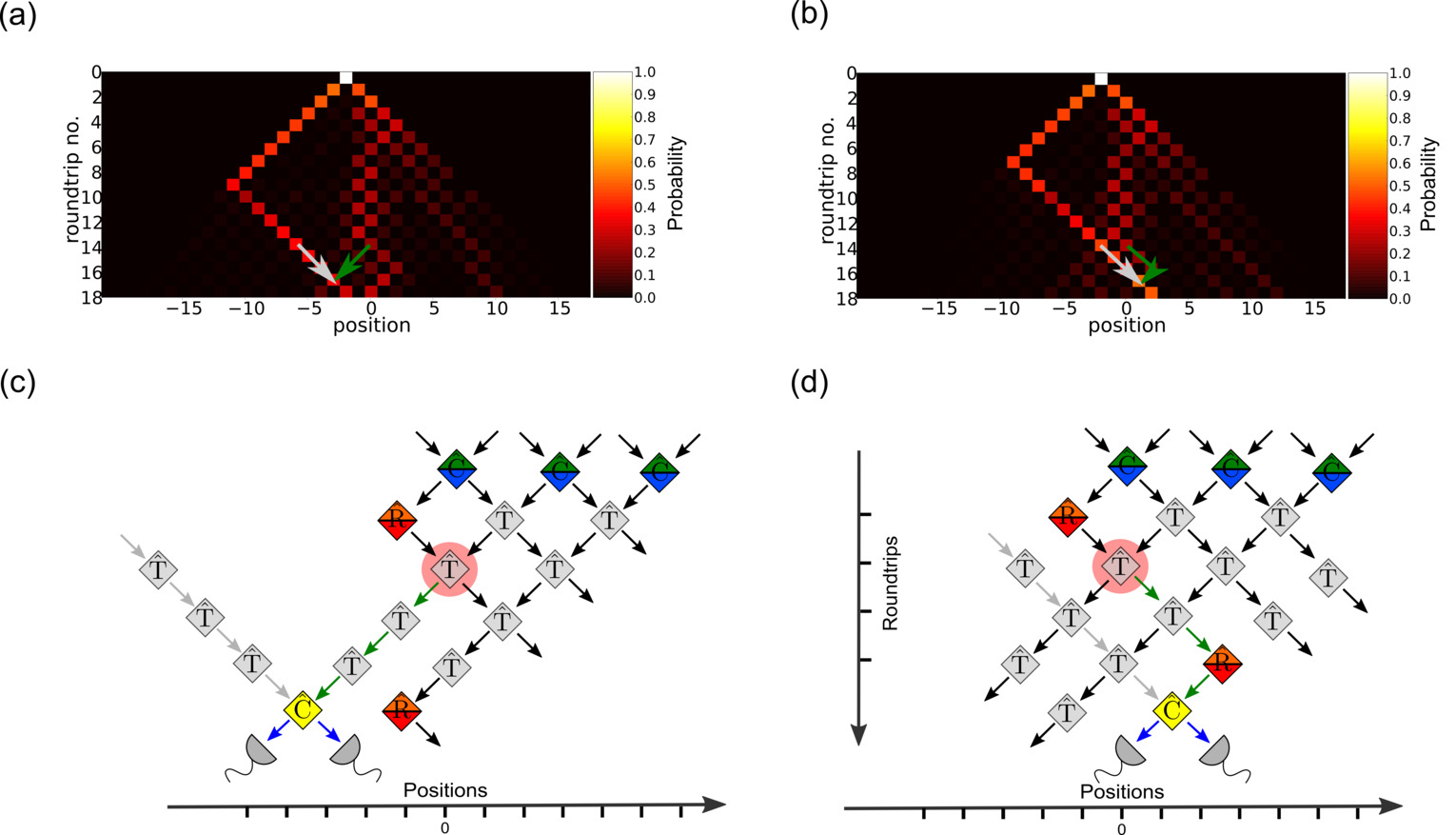}
	\caption{Plot of the intensity (polarisation is traced out) evolution for 6 steps (12 roundtrips) of a split-step quantum walk and the subsequent read-out of the vertical (a) resp. the horizontal component (b) at position 0. Subfigures (c) and (d) illustrate the corresponding read-out schemes: Either the vertical (c) or the horizontal (d) component (marked by green arrows) from the read-out position (marked by red dot) is brought to interference with the reference (grey arrows). Transmission and reflection operations are conducted such that light coming from other positions or polarisations (black arrows) does not end up in the time bins reserved for analysing the interference between reference and read-out component (marked by detector symbols). Note that the exact position of the interference is chosen such that it allows for separating one component from the others and for having the read-out in the same step for both polarisations.}
	\label{fig:Evolution}
\end{figure}

As a certain proportion of the intensity is outcoupled in each roundtrip, we are able to monitor its temporal evolution. Figure \ref{fig:Evolution} shows the evolution for 6 steps of a split-step quantum walk and the subsequent read-out of the vertical (a) resp. the horizontal component (b) at position 0. Note that in this nomenclature 0 corresponds to the innermost position of the eigenstate and not to the position at which the light pulse starts (see Figure \ref{fig:Scheme}).

\begin{figure}[t]
	\centering 
	\includegraphics[width=\columnwidth]{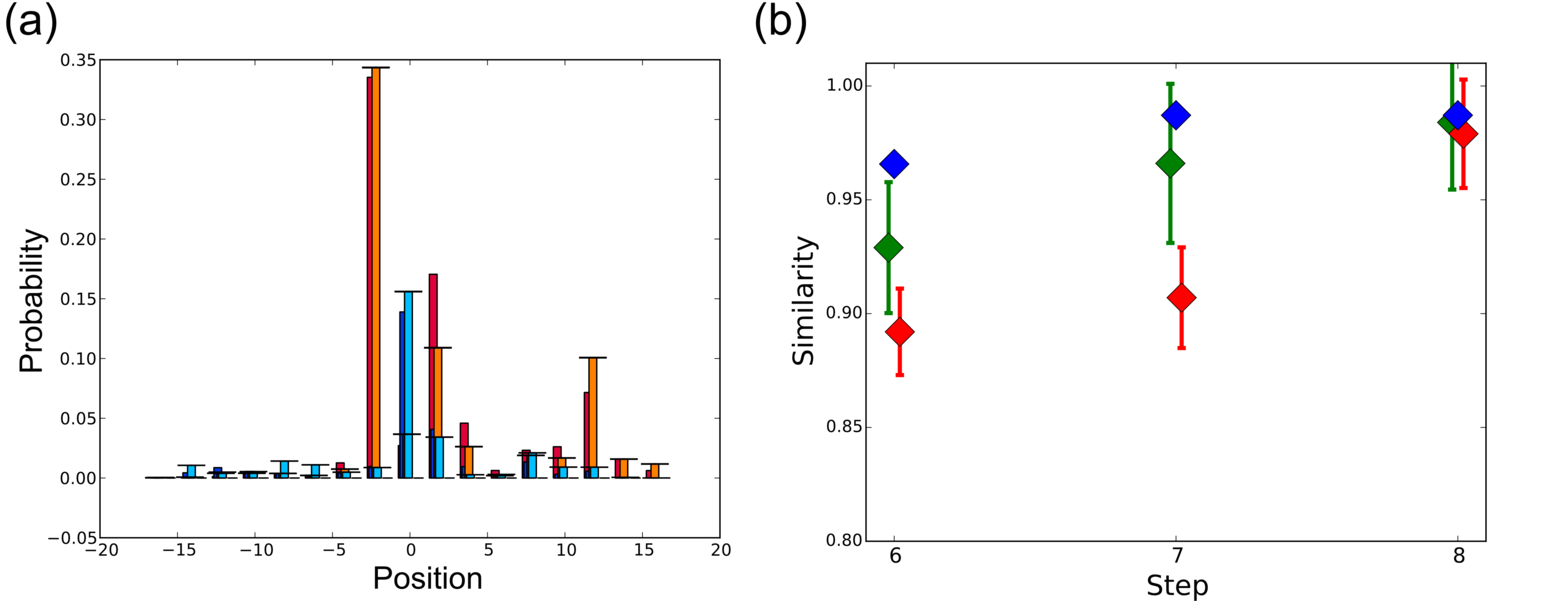}
	\caption{(a): Probability distribution in step 8 (roundtrip 17) of the split-step quantum walk. Orange (red) bar charts represent the experimental (numerical) probabilities for horizontal light, while light blue (dark blue) bar charts corresponds to experimental (numerical) data for vertical light. The large horizontal bar at positon -2 stands for the phase reference, while the components around position 10 are those not overlaping with the eigenstate.
	(b): The similarity according to \eqref{eq:similarity} between the ideal eigenstate and the numerically determined state for finite step numbers (blue markers, the same for both eigenvalues) as well as the similarity between the ideal state and the experimentally obtained state for both the regimes with eigenvalue +1 (green markers) and eigenvalue -1 (red markers.) Note that all markers correspond to integer step numbers, even though they are slightly shifted on the horizontal axis for better readability.}
	\label{fig:Similarities}
\end{figure}

The plots illustrate how the reference is steered with transmission and reflection operations on the left hand side (marked by a grey arrow) and is then brought to interference with either the vertical or the horizontal component (green arrow) of the walker evolving on the right hand side. From the distribution of intensity between the two detection bins the M-parameter can be inferred according to eq. \ref{eq:M}. During the split-step quantum walk up to roundtrip 13, the intensity on the right hand side either concentrates near position 0 or runs out towards the right. In order to quantify the overlap of this intensity near position 0 with the theoretically expected eigenstate, we calculate the similarity according to \eqref{eq:similarity} for the three innermost positions (see Figure \ref{fig:Similarities}).

While in step 6 the similarity between experiment and the ideal eigenfunction exhibits values of 0.891 $\pm$ 0.019 (eigenvalue -1) resp. 0.929 $\pm$ 0.029 (eigenvalue +1), in step 8 these values attain 0.979 $\pm$ 0.024 (eigenvalue -1) resp. 0.984 $\pm$ 0.029 (eigenvalue +1), getting almost as close to the ideal value of 1 as the numerically predicted state. The numerical simulation accounts for the limited number of steps but no other experimental imperfections, it thus quantifies the effects of the finite system size in correspondence to the deviation from the ideal similarity of 1. The difference between the numerically and the experimentally determined similarity is accordingly due to further experimental imperfections such as slightly inhomogeneous losses or imperfect EOM switchings. 
The high values of the similarity give evidence for the successful outcome of the distillation process.

\subsection{Eigenvalues}
\label{subsec:exp_eigenvalues}

Having quantified the overlap of the measured intensities with the ideal eigenstates, the focus now shifts on measuring the eigenvalues of the walk operator. We therefore monitor the evolution of the M-parameter (see \eqref{eq:M}) from step 6 to step 8 for a rotation angle $\theta_\seti = 1/4 \pi$, for which theory predicts an eigenvalue $\lambda = -1$, and $\theta_\setii = 3/4 \pi$, associated in theory with an eigenvalue of +1. This analysis is done for both horizontal and vertical polarisation, which require separate read-out procedures as explained above.

\begin{figure}[t]
	\centering 
	\includegraphics[width=\columnwidth]{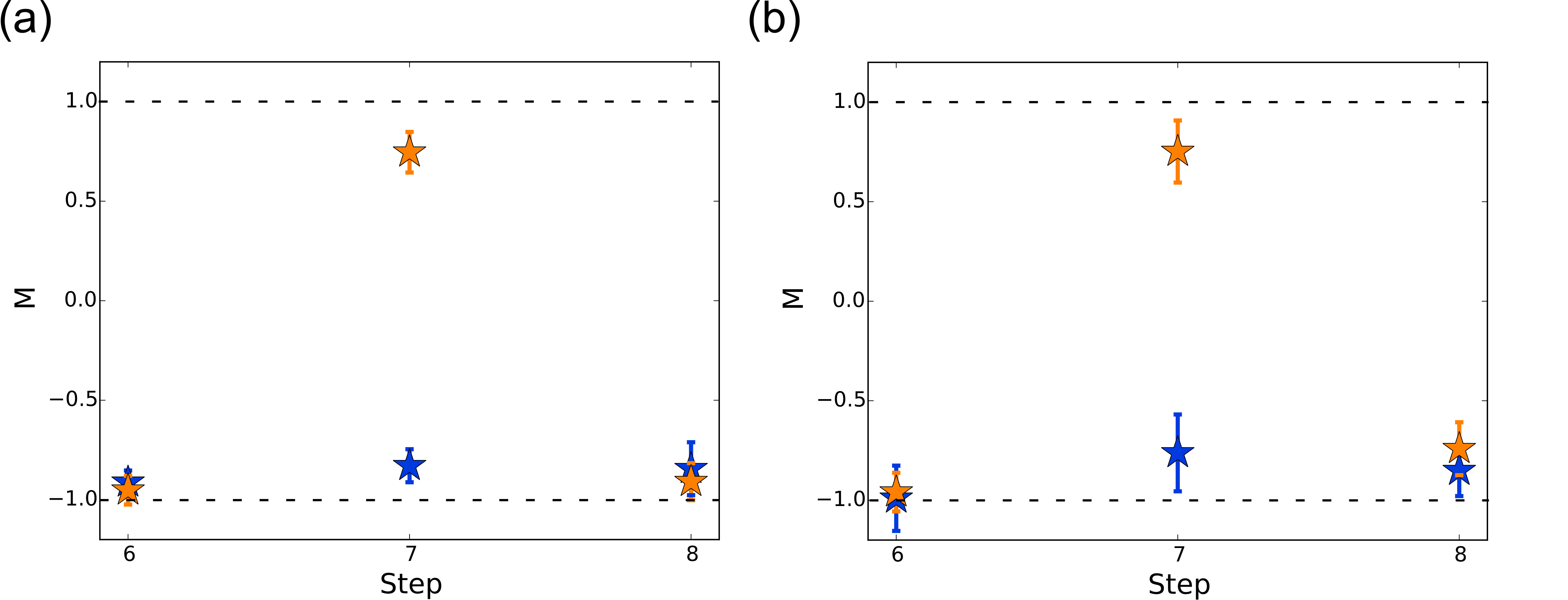}
	\caption{The evolution of the M-parameter (see \eqref{eq:M}) for position 0 from step 6 to 8 for both horizontally (a) and vertically (b) polarized light. The orange markers correspond to setting $\seti$ with Eigenvalue -1 and the blue markers to setting $\setii$ with Eigenvalue 1. The error bars are obtained in a Monte-Carlo-Simulation accounting for the effects of an error of the coupling efficiency of 2 $\%$ and an error of the coin angle of 2$^{\circ}$ (discussed in section \ref{subsec:error_discussion})}.
	\label{fig:M_1_H+V}
\end{figure}

Figure \ref{fig:M_1_H+V} shows the evolution of the M-parameter (see \eqref{eq:M}) for horizontal (a) and vertical (b) polarisation at position 0. Setting $\seti$ with Eigenvalue -1 (orange markers) and setting $\setii$ with Eigenvalue +1 (blue markers) are clearly distinguishable for both polarisations. However, the results for vertical polarisation exhibit larger error bars as we measure a more unbalanced ratio of the intensity of the walker and the intensity of the reference. This is due to the fact that the theoretical eigenstate as well as its experimental approximation show significantly less intensity in vertical polarisation than in horizontal polarisation, while the reference remains the same in both cases.

\begin{figure}[t]
	\centering 
	\includegraphics[width=\columnwidth]{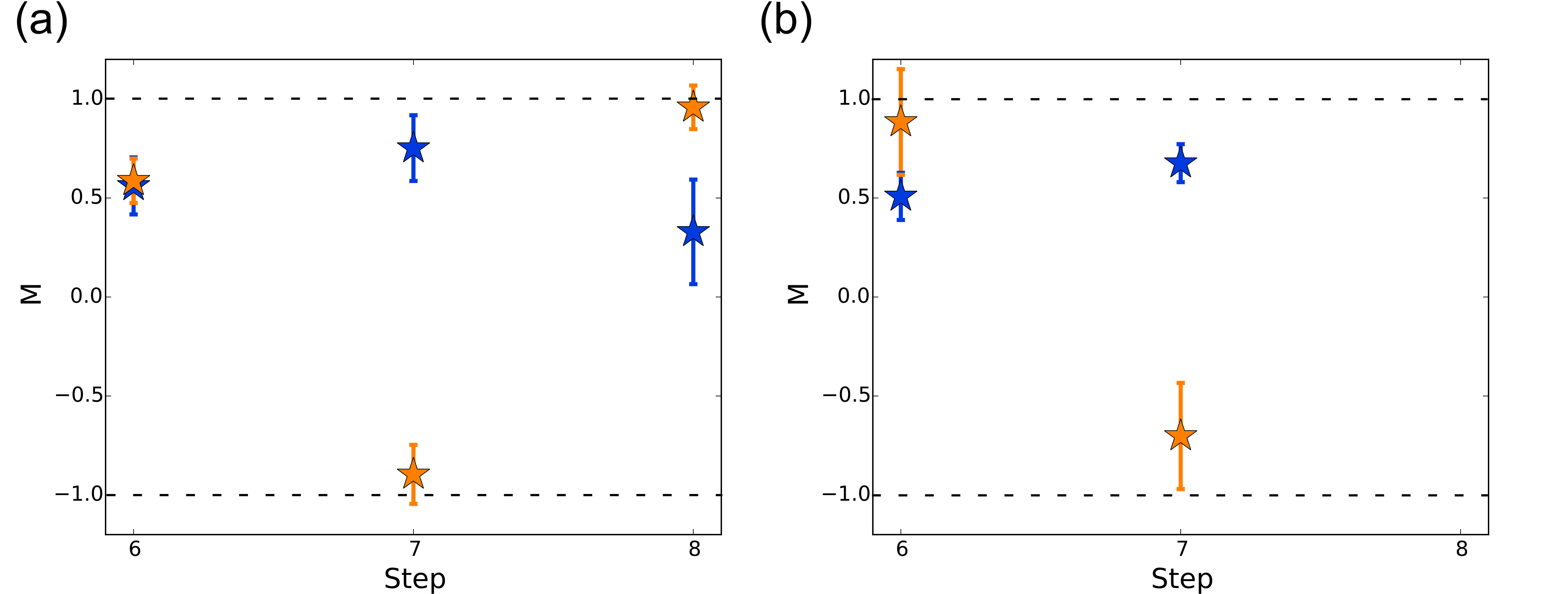}
	\caption{The evolution of the M-parameter as in Figure \ref{fig:M_1_H+V}, except now  for position 1. Note that the amplitude scaling of the eigenstate does not allow for the read-out of the vertical component in step 8.}
	\label{fig:M_2_H+V}
\end{figure}

The amplitudes of the eigenstates decrease by a factor of ($1-\sqrt{2}$) when the position is increased by one (see \eqref{eq:theo_eigenfunction}). Accordingly, the errors get bigger when reading out position 1 (see Figure \ref{fig:M_2_H+V}). For horizontal polarisation (Fig. \ref{fig:M_2_H+V} (a)) the results still reflect the theoretically expected behaviour with larger error bars than for position 0, while for vertical polarisation a quantification of the M-parameter is no longer possible in the eighth step, as the measured intensity of the walker is not significantly above the noise floor.

\subsection{Error discussion} \label{subsec:error_discussion}

The eigenstate validation as well as the read-out of the M-parameter require the measurement of intensity distributions. These distributions are subjected to inhomogeneities of the coupling efficiencies and inaccuracies in the angles of the statically and dynamically implemented coins. Assuming errors of the coupling efficiencies of 2 $\%$ and of the coin angle of 2$^{\circ}$, we conduct a Monte-Carlo simulation in which we randomly generate 1000 different settings for these quantities within the assumed error range. For each of these settings we calculate the deviation of the resulting numeric intensity distribution from a reference intensity distribution. This is obtained when running the numerical simulation with the fit parameters allowing for the closest approximation of the experimental results. The error for the individual positions and polarisations is then calculated as the standard deviation of the randomly generated samples from the reference distribution. Eventually, the errors of the similarity and the M-parameter are determined via error propagation from the errors of the intensities, resulting in the error bars visible in Figures \ref{fig:Similarities}, \ref{fig:M_1_H+V} and \ref{fig:M_2_H+V}.\\
As discussed in section \ref{subsec:state_distillation}, the eigenstates cannot be directly excited, but only be approximated via distillation during the evolution of the walker. The resulting deviations from the ideal eigenstates are quantified via the experimentally obtained similarities which are discussed in section \ref{subsec:Wavefunction}. 

\section{Conclusion} \label{sec:conclusion}

We implemented a split-step quantum walk with decoupling over step numbers ranging from 6 to 8, requiring up to 22 roundtrips in the setup. At the end of the evolution, the states in the experiment approximate the localised eigenstates with similarities of up to 0.979 $\pm$ 0.024 resp. 0.984 $\pm$ 0.029. In addition, the phase-reference method allows for measuring eigenvalues, clearly revealing two regimes with different eigenvalues for position 0. At position 1, lower intensities make the read-out more challenging, but still allow to clearly see the sign flip.
The application of the phase-reference method relies on the correct implementation of the decoupling. Our setup does not only allow for decoupling, but also serves to demonstrate that the actual decoupling coin affects the eigenvalue.
In our setup the possibilities to read out the internal degree as well as to dynamically apply different coins to different positions thus allow for investigating new aspects of topological quantum walks.
As the eigenvalues of approximate eigenstates could not be measured in previous experiments, our experiment significantly extends the range of accessible topological signatures.

\section{Acknowledgements} 

The group at Paderborn acknowledges financial support from the Gottfried Wilhelm Leibniz-Preis (grant number SI1115/3-1) and from the European Commission with the ERC project QuPoPCoRN (no.\ 725366).

The group at Hannover acknowledges support from the DFG SFB 1227 DQmat and from the ERC grant DQSIM.

C. Cedzich acknowledges support from the Excellence Initiative of the German Federal and State Governments (ZUK 81) and the DFG (project B01 of CRC 183).

\bibliographystyle{abbrvArXiv}
\bibliography{Measuring_Eigenvalues_Bib_arXiv}

\end{document}